\documentclass[sigconf]{acmart}

\copyrightyear{2020}
\acmYear{2020}
\setcopyright{acmcopyright}\acmConference[KDD '20]{Proceedings of the 26th ACM
SIGKDD Conference on Knowledge Discovery and Data Mining}{August 23--27,
2020}{Virtual Event, CA, USA}
\acmBooktitle{Proceedings of the 26th ACM SIGKDD Conference on Knowledge
Discovery and Data Mining (KDD '20), August 23--27, 2020, Virtual Event, CA, USA}
\acmPrice{15.00}
\acmDOI{10.1145/3394486.3403331}
\acmISBN{978-1-4503-7998-4/20/08}

\usepackage[ruled,vlined,boxed,linesnumbered]{algorithm2e}
\usepackage{multirow}
\usepackage{enumitem}

\begin{document}
\fancyhead{}
\title{LRSpeech: Extremely Low-Resource Speech Synthesis and Recognition}


\author{Jin Xu$^{1}$, Xu Tan$^{2}$, Yi Ren$^{3}$, Tao Qin$^{2}$, Jian Li$^{1}$, Sheng Zhao$^{4}$, Tie-Yan Liu$^{2}$}\thanks{$^{1}$This work was conducted at Microsoft.  Correspondence to: Tao Qin <taoqin@microsoft.com>.}
\affiliation{$^{1}$Institute for Interdisciplinary Information Sciences, Tsinghua University, China, $^{3}$Zhejiang University, China}
\affiliation{$^{2}$Microsoft Research Asia, $^{4}$Microsoft Azure Speech}
\email{j-xu18@mails.tsinghua.edu.cn, xuta@microsoft.com, rayeren@zju.edu.cn}
\email{lijian83@mail.tsinghua.edu.cn, {taoqin,sheng.zhao,tyliu}@microsoft.com}

\renewcommand{\shortauthors}{Xu and Tan, et al.}

\begin{abstract}
Speech synthesis (text to speech, TTS) and recognition (automatic speech recognition, ASR) are important speech tasks, and require a large amount of text and speech pairs for model training. However, there are more than 6,000 languages in the world and most languages are lack of speech training data, which
poses significant challenges when building TTS and ASR systems for extremely low-resource languages. In this paper, we develop LRSpeech, a TTS and ASR system under the extremely low-resource setting, which can support rare languages with low data cost. LRSpeech consists of three key techniques: 1) pre-training on rich-resource languages and fine-tuning on low-resource languages; 2) dual transformation between TTS and ASR to iteratively boost the accuracy of each other; 3) knowledge distillation to customize the TTS model on a high-quality target-speaker voice and improve the ASR model on multiple voices. We conduct experiments on an experimental language (English) and a truly low-resource language (Lithuanian) to verify the effectiveness of LRSpeech. Experimental results show that LRSpeech 1) achieves high quality for TTS in terms of both intelligibility (more than $98\%$ intelligibility rate) and naturalness (above 3.5 mean opinion score (MOS)) of the synthesized speech, which satisfy the requirements for industrial deployment, 2) achieves promising recognition accuracy for ASR, and 3) last but not least, uses extremely low-resource training data. We also conduct comprehensive analyses on LRSpeech with different amounts of data resources, and provide valuable insights and guidances for industrial deployment. We are currently deploying LRSpeech into a commercialized cloud speech service to support TTS on more rare languages. 

\end{abstract}

\maketitle

\section{Introduction}
Speech synthesis (text to speech, TTS)~\cite{wang2017tacotron,shen2018natural,ping2018deep,ren2019fastspeech} and speech recognition (automatic speech recognition, ASR)~\cite{chorowski2014end,chan2016listen,chiu2018state} are two key tasks in speech domain, and attract a lot of attention in both the research and industry community. However, popular commercialized speech services (e.g., Microsoft Azure, Google Cloud, Nuance, etc.) only support dozens of languages for TTS and ASR, while there are more than 6,000 languages in the world~\citep{lewis2013simons}. Most languages are lack of speech training data, which makes it difficult to support TTS and ASR for these rare languages, as large-amount and high-cost speech training data are required to ensure good accuracy for industrial deployment.

We describe the typical training data to build TTS and ASR systems as follows:
\begin{itemize}[leftmargin=*]
\item TTS aims to synthesize intelligible and natural speech from text sequences, and usually needs single-speaker high-quality recordings that are collected in professional recording studio. To improve the pronunciation accuracy, TTS also requires a pronunciation lexicon to convert the character sequence into phoneme sequence as the model input (e.g., ``speech" is converted into ``s p iy ch"), which is called as grapheme-to-phoneme conversion~\citep{sun2019token}. Additionally, TTS models use text normalization rules to convert the irregular word into the normalized type that is easier to pronounce (e.g., ``Sep 7th" is converted into ``September seventh"). 
\item ASR aims to generate correct transcripts (text) from speech sequences, and usually requires speech data from multiple speakers in order to generalize to unseen speakers during inference. The multi-speaker speech data in ASR do not need to be as high-quality as that in TTS, but the data amount is usually an order of magnitude bigger. We call the speech data for ASR as multi-speaker low-quality data\footnote{The low quality here does not mean the quality of ASR data is very bad, but is just relatively low compared to the high-quality TTS recordings.}. Optionally, ASR can first recognize the speech into phoneme sequence, and further convert it into character sequence with the pronunciation lexicon as in TTS.
\item Besides paired speech and text data, TTS and ASR models can also leverage unpaired speech and text data to further improve the performance. 
\end{itemize}

\begin{table*}[h]
\small
\centering
\begin{tabular}{l l | c | c | c | c }
\toprule
\multicolumn{2}{l|}{Setting} & Rich-Resource & Low-Resource & Extremely Low-Resource & Unsupervised \\
\midrule
\multirow{6}{*}{Data} & pronunciation lexicon & $\checkmark$ & $\checkmark$ & $\times$ & $\times$ \\
& paired data (single-speaker, high-quality) & dozens of hours & dozens of minutes & several minutes & $\times$ \\
& paired data (multi-speaker, low-quality) & hundreds of hours & dozens of hours & several hours & $\times$ \\
& unpaired speech (single-speaker, high-quality) & $\checkmark$ & dozens of hours & $\times$ & $\times$ \\
& unpaired speech (multi-speaker, low-quality) & $\checkmark$ & $\checkmark$ & dozens of hours & $\checkmark$\\
& unpaired text & $\checkmark$ & $\checkmark$ & $\checkmark$ & $\checkmark$ \\
\midrule
\multirow{2}{*}{Related Work} & TTS & \cite{shen2018natural,ping2018deep,li2019neural,ren2019fastspeech} & \cite{baevski2019effectiveness,chung2019semi,liu2019towards,ren2019almost} & \multirow{2}{*}{Our Work} & / \\
\cmidrule{2-4} \cmidrule{6-6}
& ASR &\cite{chorowski2014end,chan2016listen,chiu2018state} & \cite{tjandra2017listening,hori2019cycle,rosenberg2019speech,schneider2019wav2vec} &  & ~\cite{yeh2018unsupervised,chen2018towards,liu2018completely} \\ 
\bottomrule
\end{tabular}
\caption{The data resource to build TTS and ASR systems and the corresponding related works in rich-resource, low-resource, extremely low-resource and unsupervised settings. }
\label{tab_resource_setting}
\vspace{-4mm}
\end{table*}

\subsection{Related Work}
According to the data resource used, previous works on TTS and ASR can be categorized into rich-resource, low-resource and unsupervised settings. 

As shown in Table~\ref{tab_resource_setting}, we list the data resources and the corresponding related works in each setting
:
\begin{itemize}[leftmargin=*]
\item In the rich-resource setting, both TTS~\cite{shen2018natural,ping2018deep,li2019neural,ren2019fastspeech} and ASR~\cite{chorowski2014end,chan2016listen,chiu2018state} require a large amount of paired speech and text data to achieve high accuracy: TTS usually needs dozens of hours of single-speaker high-quality recordings, while ASR requires at least hundreds of hours multiple-speaker low-quality data. Besides, TTS in the rich-resource setting also leverages pronunciation lexicon for accurate pronunciation. Optionally, unpaired speech and text data can be leveraged. 
\item In the low-resource setting, the single-speaker high-quality paired data are reduced to dozens of minutes in TTS~\cite{baevski2019effectiveness,chung2019semi,liu2019towards,ren2019almost} while the multi-speaker low-quality paired data is reduced to dozens of hours in ASR~\cite{tjandra2017listening,hori2019cycle,rosenberg2019speech,schneider2019wav2vec}, compared to that in the rich-resource setting. Additionally, they leverage unpaired speech and text data to ensure the performance.
\item In the unsupervised setting, only unpaired speech and text data are leverage to build ASR models~\cite{yeh2018unsupervised,chen2018towards,liu2018completely}.
\end{itemize}

As can be seen, a large amount of data resources are leveraged in the rich-resource setting to ensure the accuracy for industrial deployment. Considering nearly all low-resource languages are lack of training data and there are more than 6,000 languages in the world, it will be a huge cost for training data collection. Although data resource can be reduced in the low-resource setting, it still requires 1) a certain amount of paired speech and text (dozens of minutes for TTS and dozens of hours for ASR), 2) a pronunciation lexicon, and 3) a large amount of single-speaker high-quality unpaired speech data that still incur high data collection cost\footnote{Although we can crawl the multi-speaker low-quality unpaired speech data from the web, it is hard to crawl the single-speaker high-quality unpaired speech data. Therefore, it has the same collection cost (recorded by human) with the single-speaker high-quality paired data.}. What is more, the accuracy of the TTS and ASR models in the low-resource setting is not high enough. The purely unsupervised methods for ASR suffer from low accuracy and cannot meet the requirement of industrial deployment.

\subsection{Our Method}
In this paper, we develop LRSpeech, a TTS and ASR system under the extremely low-resource setting, which supports rare languages with low data collection cost. LRSpeech aims for industrial deployment under two constraints: 1) extremely low data collection cost, and 2) high accuracy to satisfy the deployment requirement. For the first constraint, as the extremely low-resource setting shown in Table~\ref{tab_resource_setting}, LRSpeech explores the limits of data requirements by 1) using single-speaker high-quality paired data as few as possible (several minutes), 2) using a few multi-speaker low-quality paired data (several hours), 3) using slightly more multi-speaker low-quality unpaired speech data (dozens of hours), 4) not using single-speaker high-quality unpaired data, and 5) not using the pronunciation lexicon but directly taking character as the input of TTS and the output of ASR. 

For the second constraint, LRSpeech leverages several key techniques including transfer learning from rich-resource languages, iterative accuracy boosting between TTS and ASR through dual transformation, and knowledge distillation to further refine TTS and ASR models for better accuracy. Specifically, LRSpeech consists of a three-stage pipeline:
\begin{itemize}[leftmargin=*]
\item We first pre-train both TTS and ASR models on rich-resource languages with plenty of paired data, which can learn the alignment capability between speech and text and benefit the alignment learning on low-resource languages. 
\item We further leverage dual transformation between TTS and ASR to iteratively boost the accuracy of each other with unpaired speech and text data.
\item Furthermore, we leverage knowledge distillation with unpaired speech and text data to customize the TTS model on a high-quality target-speaker voice and improve the ASR model on multiple voices.
\end{itemize}


\begin{figure*}[!t]
\centering
\includegraphics[width=1.0\textwidth]{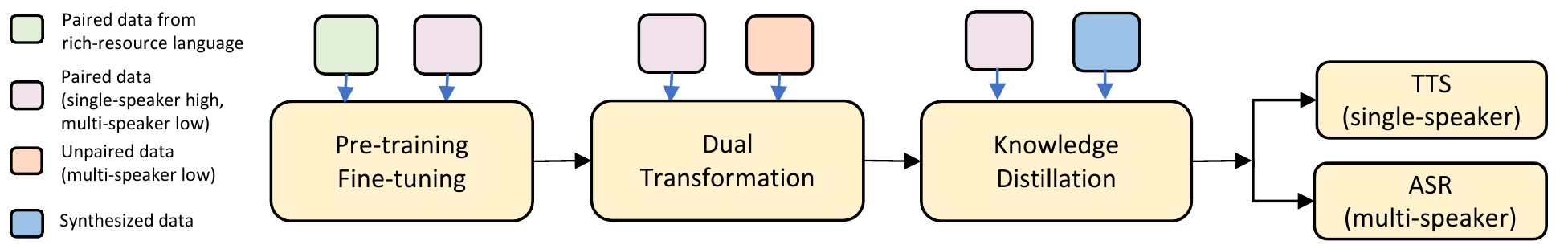}
\caption{The three-stage pipeline of LRSpeech.}
\label{fig_pipline}
\vspace{-2mm}
\end{figure*}

\subsection{Data Cost and Accuracy}
Next, we introduce the extremely low data cost while promising accuracy achieved by LRSpeech. 

According to~\cite{yamagishi2010thousands,harband_2010,thu2016comparison,bruguier2018dictionary,cooper2019text}, the pronunciation lexicon, single-speaker high-quality paired data and single-speaker high-quality unpaired speech data require much higher collection cost than other data such as multi-speaker low-quality unpaired speech data and unpaired text, since they can be crawled from the web. Accordingly, compared to the low-resource setting in Table~\ref{tab_resource_setting}, LRSpeech 1) removes the pronunciation lexicon, 2) reduces the single-speaker high-quality paired data by an order of magnitude, 3) removes single-speaker high-quality unpaired speech data, 4) also reduces multi-speaker low-quality paired data by an order of magnitude, 5) similarly leverages multi-speaker low-quality unpaired speech, and 6) additionally leverage paired data from rich-resource languages which incur no additional cost since they are already available in the commercialized speech service. Therefore, LRSpeech can greatly reduce the data collection cost for TTS and ASR.

To verify the effectiveness of LRSpeech under the extremely low-resource setting, we first conduct comprehensive experimental studies on English and then verify on the truly low-resource language: Lithuanian, which is for product deployment. For TTS, LRSpeech achieves 98.08\% intelligibility rate, 3.57 MOS score, with 0.48 gap to the ground-truth recordings, satisfying the online deployment requirements\footnote{According to the requirements of a commercialized cloud speech service, the intelligibility rate should be higher than 98\% and the MOS score should be higher than 3.5 while the MOS gap to the ground-truth recordings should be less than 0.5.}. For ASR, LRSpeech achieves 28.82\% WER and 14.65\% CER, demonstrating great potential under the extremely low-resource setting. Furthermore, we also conduct ablation studies to verify the effectiveness of each component in LRSpeech, and analyze the accuracy of LRSpeech under different data settings, which provide valuable insights for industrial deployment. Finally, we apply LRSpeech to Lithuanian and also meets the online requirement for TTS and achieves promising results on ASR. We are currently deploying LRSpeech to a commercialized speech service to support TTS for rare languages.

\section{LRSpeech}

In this section, we introduce the details of LRSpeech for extremely low-resource speech synthesis and recognition. We first give an overview of LRSpeech, and then introduce the formulation of TTS and ASR. We further introduce each component of LRSpeech respectively, and finally describe the model structure of LRSpeech. 

\subsection{Pipeline Overview}
To ensure the accuracy of TTS and ASR models under extremely low-resource scenarios, we design a three-stage pipeline for LRSpeech as shown in Figure~\ref{fig_pipline}:
\begin{itemize}[leftmargin=*]
\item Pre-training and fine-tuning. We pre-train both TTS and ASR models on rich-resource languages and then fine-tune them on low-resource languages. Leveraging rich-resource languages in LRSpeech are based on two considerations: 1) a large amount of paired data on rich-resource languages are already available in the commercialized speech service, and 2) the alignment capability between speech and text in rich-resource languages can benefit the alignment learning in low-resource languages, due to the pronunciation similarity between human languages~\cite{wind1989evolutionary}.
\item Dual transformation. Considering the dual nature between TTS and ASR, we further leverage dual transformation~\citep{ren2019almost} to boost the accuracy of each other with unpaired speech and text data.
\item Knowledge distillation. To further improve the accuracy of TTS and ASR and facilitate online deployment, we leverage knowledge distillation~\citep{kim2016sequence,tan2018multilingual} to synthesize paired data to train better TTS and ASR models.
\end{itemize}

\subsection{Formulation of TTS and ASR}
TTS and ASR are usually formulated as sequence to sequence problems~\citep{wang2017tacotron,chan2016listen}. Denote the text and speech sequence pair $(x,y) \in D$, where $D$ is the paired text and speech corpus. Each element in the text sequence $x$ represents a phoneme or character, while each element in the speech sequence $y$ represents a frame of speech. To learn the TTS model $\theta$, a mean square error loss is used:
\begin{equation}
\begin{aligned}
\mathcal{L}(\theta; D) = -\Sigma_{(x,y)\in D} (y - f(x;\theta))^2.
\end{aligned}
\label{eq_tts_loss}
\end{equation}
To learn the ASR model $\phi$, a negative log likelihood loss is used:
\begin{equation}
\begin{aligned}
\mathcal{L}(\phi; D) = -\Sigma_{(y,x)\in D}\log P(x|y; \phi).  
\end{aligned}
\label{eq_asr_loss}
\end{equation}
TTS and ASR models can be developed based on an encoder-attention-decoder framework~\cite{bahdanau2014neural,luong2015effective,vaswani2017attention}, where the encoder transforms the source sequence into a set of hidden representations, and the decoder generates the target sequence autoregressively based on the source hidden representations obtained through an attention mechanism~\citep{bahdanau2014neural}.

We make some notations for the data used in LRSpeech. Denote $D_{\text{rich\_tts}}$ as the high-quality TTS paired data in rich-resource languages, $D_{\text{rich\_asr}}$ as the low-quality ASR paired data in rich-resource languages, $D_h$ as the single-speaker high-quality paired data for target speaker, and $D_l$ as the multi-speaker low-quality paired data. Denote $X^u$ as unpaired text data while $Y^u$ as multi-speaker low-quality unpaired speech data.

Next, we introduce each component of the LRSpeech pipeline in the following subsections.

\subsection{Pre-Training and Fine-Tuning}
The key to the conversion between text and speech is to learn the alignment between the character/phoneme representations (text) and the acoustic features (speech). Since people coming from different nations and speaking different languages share similar vocal organs and thus similar pronunciations, the ability of alignment learning in one language can help the alignment in another language~\citep{wind1989evolutionary,kuhl2008phonetic}. This motivates us to transfer the TTS and ASR models that trained in rich-resource languages into low-resource languages, considering there are plenty of paired speech and text data for both TTS and ASR in rich-resource languages.

\paragraph{Pre-Training} 
We pre-train the TTS model $\theta$ with data corpus $D_{\text{rich\_tts}}$ following Equation~\ref{eq_tts_loss} and pre-train the ASR model $\phi$ with $D_{\text{rich\_asr}}$ following Equation~\ref{eq_asr_loss}.

\paragraph{Fine-Tuning} 
Considering the rich-resource and low-resource languages have different phoneme/character vocabularies and speakers, we initialize the TTS and ASR models on low-resource language with all the pre-trained parameters except the phoneme/character and speaker embeddings in TTS and the phoneme/character embeddings in ASR\footnote{ASR model does not need speaker embeddings, and the target embeddings and the softmax matrix are usually shared in many sequence generation tasks for better accuracy~\cite{press2017using}.} respectively. We then fine-tune the TTS model $\theta$ and ASR model $\phi$ both with the concatenation corpus of $D_{h}$ and $D_{l}$ following Equation~\ref{eq_tts_loss} and Equation~\ref{eq_asr_loss} respectively.
During fine-tuning, we first fine-tune the character embeddings and speaker embeddings following the practice in~\cite{chen2019extensible,artetxe2019cross}, and then fine-tune all parameters. It can help prevent the TTS and ASR models from overfitting on the limited paired data in a low-resource language.


\subsection{Dual Transformation between TTS and ASR} \label{sec_dt_unseen}
TTS and ASR are two dual tasks and their dual nature can be explored to boost the accuracy of each other, especially in the low-resource scenarios. Therefore, we leverage dual transformation~\cite{ren2019almost} between TTS and ASR to improve the ability to transform between text and speech. Dual transformation shares similar ideas with back-translation~\cite{sennrich2016improving} in machine translation and cycle-consistency~\cite{zhu2017unpaired} in image translation, which are effective ways to leverage unlabeled data in speech, text and image domains respectively. 
Dual transformation works as follows:
\begin{itemize}[leftmargin=*]
\item For each unpaired text sequence $x \in X^u$, we transform it into speech sequence using the TTS model $\theta$, and construct a pseudo corpus $D(X^u)$ to train the ASR model $\phi$ following Equation~\ref{eq_asr_loss}.
\item For each unpaired speech sequence $y \in Y^u$, we transform it into text sequence using the ASR model $\phi$, and construct a pseudo corpus $D(Y^u)$ to train the TTS model $\theta$ following Equation~\ref{eq_tts_loss}.  
\end{itemize}

During training, we run the dual transformation process on the fly, which means the pseudo corpus are updated in each iteration and the model can benefit from the newest data generated by each other. Next, we introduce some specific designs in dual transformation to support multi-speaker TTS and ASR.

\paragraph{Multi-Speaker TTS Synthesis} 
Different from~\cite{ren2019almost,liu2019towards} that only support a single speaker in both TTS and ASR model, we support multi-speaker TTS and ASR in the dual transformation stage. Specifically, we randomly choose a speaker ID and synthesize speech of this speaker given a text sequence, which can benefit the training of the multi-speaker ASR model. Furthermore, the ASR model transforms multi-speaker speech into text, which can help the training of the multi-speaker TTS model.

\paragraph{Levering Unpaired Speech of Unseen Speakers}
Since multiple-speaker low-quality unpaired speech data are much easier to obtain than high-quality single-speaker unpaired speech data, enabling the TTS and ASR models to utilize unseen speakers' unpaired speech in dual transformation can make our system more robust and scalable. Compared to ASR, it is more challenging for TTS to synthesize voice on unseen speakers. To this end, we split dual transformation into two phases: 1) In the first phase, we only use the unpaired speech whose speakers are seen before in the training data. 2) In the second phase, we also add the unpaired speech whose speakers are unseen in the training data. As the ASR model can naturally support unseen speakers, the pseudo paired data can be used to train and enable the TTS model with the capability to synthesize speech of new speakers. 


\subsection{Customization on TTS and ASR through Knowledge Distillation}
The TTS and ASR models we currently have are far from ready for online deployment after dual transformation. There are several issues we need to address: 1) While the TTS model can support multiple speakers, the speech quality of our target speaker is not good enough and needs further improvement; 2) The synthesized speech by the TTS models still have word skipping and repeating issues; 3) The accuracy of the ASR model needs to be further improved. Therefore, we further leverage knowledge distillation~\cite{kim2016sequence,tan2018multilingual}, which generates target sequences given source sequences as input to construct a pseudo corpus, to customize the TTS and ASR models for better accuracy. 

\begin{figure*}[t] 
\centering
\includegraphics[width=\textwidth]{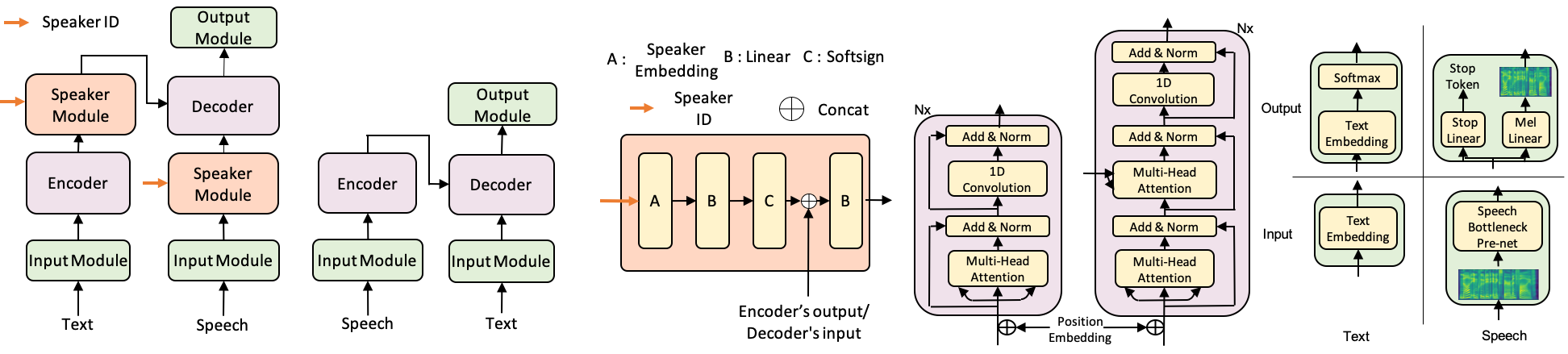}
\begin{tabular}{lllll}
  \hspace{0.7cm}(a) TTS model  &
  \hspace{1.1cm}(b) ASR model & 
  \hspace{0.8cm}(c) Speaker module &
  \hspace{1.0cm}(d) Encoder (left) and & 
  \hspace{0.6cm}(e) Input/Output module 
  \\
  \hspace{5cm}& & & \hspace{1.5cm} Decoder (right) & \hspace{1.1cm} for speech/text
\end{tabular}
\vspace{-1mm}
\caption{The Transformer based TTS and ASR models in LRSpeech.}
\label{fig_TTS_ASR_model}
\vspace{-2mm}
\end{figure*}

\subsubsection{Knowledge Distillation for TTS} \label{method:Distill:TTS}
 The knowledge distillation process for TTS consists of three steps: 
\begin{itemize}[leftmargin=*]
    \item For each unpaired text sequence $x \in X^u$, we synthesize the corresponding speech of the target speaker using the TTS model $\theta$, and construct a single-speaker pseudo corpus $D(X^u)$.
    \item Filter the pseudo corpus $D(X^u)$ whose synthesized speech has word skipping and repeating issues.
    \item Use the filtered corpus $D(X^u)$ to train a new TTS model dedicated to the target speaker following Equation~\ref{eq_tts_loss}. 
\end{itemize}

In the first step, the speech in the pseudo corpus $D(X^u)$ are single-speaker, which is different from the multi-speaker pseudo corpus $D(X^u)$ in Section~\ref{sec_dt_unseen}. The TTS model (obtained by dual transformation) in the first step has word skipping and repeating issues. Therefore, in the second step, we filter the synthesized speech which has word skipping and repeating issues, and thus the distilled model can be trained on accurate text and speech pairs. In this way, the word skipping and repeating problem can be largely reduced. We filter the synthesized speech based on two metrics: word coverage ratio (WCR) and attention diagonal ratio (ADR).

\paragraph{Word Coverage Ratio} We observe that word skipping happens when a word has small or no attention weights from the target mel-spectrograms. Therefore, we propose word coverage ratio (WCR):
\begin{equation}
WCR =\min_{i \in [1, N]} \{\max_{t \in [1, T_i]} \max_{s \in [1, S]} A_{t,s} \},
\end{equation}
where $N$ is the number of words in a sentence, $T_i$ is the number of characters in the $i$-th word, $S$ is the number of frames of the target mel-spectrograms, and $A_{t,s}$ denotes the element in the $t$-th row and $s$-th column of the attention weight matrix $A$. We get the attention weight matrix $A$ from the encoder-decoder attention weights in the TTS model and calculate the mean over different layers and attention heads. A high WCR indicates all words in a sentence have high attention weights from target speech frames, and thus is less likely to cause word skipping.

\paragraph{Attention Diagonal Ratio} As demonstrated by previous works~\citep{ren2019fastspeech,wang2017tacotron}, the attention alignments between text and speech are monotonic and diagonal. When the synthesized speech has word skipping and repeating issues, or is totally crashed, the attention alignments will deviate from the diagonal. We define the attention diagonal ratio (ADR) as:
\begin{equation}
ADR=\frac{\sum_{t=1}^{T} \sum_{s=kt-b}^{kt+b} A_{t,s}}{\sum_{t=1}^{T} \sum_{s=1}^{S} A_{t,s}},
\end{equation}
where $T$ and $S$ are the number of characters and speech frames in a text and speech pair, $k=\frac{S}{T}$, and $b$ is a hyperparameter to determine the width of diagonal. ADR measures how much attention lies in the diagonal area with a width of $b$. A higher ADR indicates that the synthesized speech has good attention alignment with text and thus has less word skipping, repeating or crashing issues.



\subsubsection{Knowledge Distillation for ASR} 
Since the unpaired text and low-quality multi-speaker unpaired speech are both available for ASR
, we leverage both the ASR and TTS models to synthesize data during the knowledge distillation for ASR:
\begin{itemize}[leftmargin=*]
    \item For each unpaired speech $y \in Y^u$, we generate the corresponding text using the ASR model $\phi$, and construct a pseudo corpus $D(Y^u)$.
    \item For each unpaired text $x \in X^u$, we synthesize the corresponding speech of multiple speakers using the TTS model $\theta$, and construct a pseudo corpus $D(X^u)$.
    \item We combine the above pseudo corpus $D(Y^u)$ and $D(X^u)$, as well as the single-speaker high-quality paired data $D_h$ and multi-speaker low-quality paired data $D_l$ to train a new ASR model following Equation~\ref{eq_asr_loss}.
\end{itemize}

Similar to the knowledge distillation for TTS, we also leverage a large amount of unpaired text to synthesize speech. To further improve the ASR accuracy, we use SpecAugment~\cite{park2019specaugment} to add noise in the input speech which acts like data augmentation.

\subsection{Model Structure of LRSpeech}
In this section, we introduce the model structure of LRSpeech, as shown in Figure~\ref{fig_TTS_ASR_model}.

\paragraph{Transformer Model}
Both the TTS and ASR models adopt the Transformer based encoder-attention-decoder structure~\cite{vaswani2017attention}. One difference from the original Transformer model is that we replace the feed-forward network with a one-dimensional convolution network following~\cite{ren2019almost}, in order to better capture the dependencies in a long speech sequence.

\paragraph{Input/Output Module}
To enable the Transformer model to support ASR and TTS, we need different input and output modules for speech and text~\citep{ren2019almost}. For the TTS model: 1) The input module of the encoder is a character/phoneme embedding lookup table, which converts character/phoneme ID into embedding; 2) The input module of the decoder is a speech pre-net, which consists of multiple dense layers to transform each speech frame non-linearly; 3) The output module of the decoder consists of a linear layer to convert hidden representations into mel-spectrograms, and a stop linear layer with a sigmoid function to predict whether current step should stop or not. For the ASR model: 1) The input module of the encoder consists of multiple convolutional layers, which reduce the length of the speech sequence; 2) The input module of the decoder is a character/phoneme embedding lookup table; 3) The output module of the decoder consists of a linear layer and a softmax function, where the linear layer shares the same weights with the character/phoneme embedding lookup table in the decoder input module.

\paragraph{Speaker Module}
The multi-speaker TTS model relies on a speaker embedding module to differentiate multiple speakers. We add a speaker embedding vector both in the encoder output and decoder input (after the decoder input module). As shown in Figure~\ref{fig_TTS_ASR_model}~(c), we convert the speaker ID into a speaker embedding vector using an embedding lookup table, and then add a linear transformation with a softsign function $x=x/(1+|x|)$. We further concatenate the obtained vector with the encoder output or decoder input, and use another linear layer to reduce the hidden dimension to the original hidden of the encoder output or decoder input.

\section{Experiments and Results}
In this section, we conduct experiments to evaluate LRSpeech for extremely low-resource TTS and ASR. We first describe the experiment settings, show the results of our method, and conduct some analyses of LRSpeech.

\subsection{Experimental Setup} 
\subsubsection{Datasets} \label{sec_exp_data}
We describe the datasets used in rich-resource and low-resource languages respectively:
\begin{itemize}[leftmargin=*]
    \item We select Mandarin Chinese as the rich-resource language. The TTS corpus $D_{\text{rich\_tts}}$ contains 10000 paired speech and text data (12 hours) of a single speaker from Data Baker\footnote{https://www.data-baker.com/open\_source.html}. The ASR corpus $D_{\text{rich\_asr}}$ is from AIShell~\cite{bu2017aishell}, which contains about 120000 paired speech and text data (178 hours) from 400 Mandarin Chinese speakers.
    \item We select English as a low-resource language for experimental development. The details of the data resources used are shown in Table~\ref{table_exp_data}. More information about these datasets are shown in Section~\ref{repro:dataset} and Table~\ref{fig_dataset}.
\end{itemize}



\begin{table}[!t]
\small
\centering
\begin{tabular}{lllll}
\toprule
Notation       & Quality & Type                   & Dataset     & \#Samples        \\\midrule
$D_h$   & High    & Paired  & LJSpeech~\citep{ljspeech17}    & 50 (5 minutes) \\
$D_l$   & Low     & Paired  & LibriSpeech~\cite{panayotov2015librispeech} & 1000 (3.5 hours)          \\
$Y^u_{\text{seen}}$ & Low     & Unpaired         & LibriSpeech & 2000 (7 hours)           \\
$Y^u_{\text{unseen}}$ & Low     & Unpaired         & LibriSpeech & 5000 (14 hours)  \\
$X^u$ & /   & Unpaired   & news-crawl  & 20000          \\
\bottomrule
\end{tabular}
\caption{The data used in the low-resource language: English. $D_h$ represents target-speaker high-quality paired data. $D_l$ represents multi-speaker low-quality paired data (50 speakers). $Y^u_{\text{seen}}$ represents multi-speaker low-quality unpaired speech data (50 speakers), where speakers are seen in the paired training data. $Y^u_{\text{unseen}}$ represents multi-speaker low-quality unpaired speech data (50 speakers), where speakers are unseen in the paired training data. $X_u$ represents unpaired text data.}
\label{table_exp_data}
\vspace{-4mm}
\end{table}


\subsubsection{Training and Evaluation} 
We use a 6-layer encoder and a 6-layer decoder for both the TTS and ASR models. The hidden size, character embedding size, and speaker embedding size are all set to 384, and the number of attention heads is set to 4. During dual transformation, we up-sample the paired data to make its size roughly the same with the unpaired data. During knowledge distillation, we filter the synthesized speech with WCR less than 0.7 and ADR less than 0.7. The width of diagonal ($b$) in ADR is 10. More model training details are introduced in Section~\ref{repro:model_config}.

The TTS model uses Parallel WaveGAN~\cite{yamamoto2019parallel} as the vocoder to synthesize speech. To train Parallel WaveGAN, we combine the speech data in the Mandarin Chinese TTS corpus $D_{\text{rich\_tts}}$ with the speech data in the English target-speaker high-quality corpus $D_h$. We up-sample the speech data in $D_h$ to make it roughly the same with the speech data in $D_{\text{rich\_tts}}$.

For evaluation, we use MOS (mean opinion score) and IR (intelligibility rate) for TTS, and WER (word error rate) and CER (character error rate) for ASR. 
For TTS, we select English text sentences from the news-crawl\footnote{http://data.statmt.org/news-crawl} dataset to synthesize speech for evaluation. We randomly select 200 sentences for IR test and 20 sentences for MOS test, following the practice in ~\citep{wang2017tacotron,ren2019fastspeech}\footnote{The sentences for IR and MOS test, audio samples and test reports can be founded in https://speechresearch.github.io/lrspeech.}. Each speech is listened by at least 5 testers for IR test and 20 testers for MOS test, who are all native English speakers. For ASR, we measure the WER and CER score on the LibriSpeech ``test-clean'' set. The test sentences and speech for TTS and ASR do not appear in the training corpus.

\subsection{Results}
\label{sec_exp_result}

\begin{table}[!h]
\centering
\begin{tabular}{l | l l | l l}
\toprule
\multirow{2}{*}{Setting} & \multicolumn{2}{c|}{TTS} & \multicolumn{2}{c}{ASR}  \\
\cmidrule{2-3} \cmidrule{4-5} 
& IR (\%) & MOS & WER (\%) & CER (\%) \\
\midrule
Baseline \#1   &  / &  /  &   148.29 & 100.16 \\
Baseline \#2   &  /  &  /  &   122.09 & 97.91  \\
\midrule
~+PF      &  93.09  &  2.84   &   103.70 & 69.53  \\
~+PF+DT &  96.70  &  3.28   &   38.94 & 19.99  \\
~+PF+DT+KD (LRSpeech) & \textbf{98.08} & \textbf{3.57} & \textbf{28.82} & \textbf{14.65} \\
\midrule
GT (Parallel WaveGAN)   &  -  &  3.88   &  -  & - \\
GT           &  -  &  4.05   &  -  & - \\
\bottomrule
\end{tabular}
\caption{The accuracy comparisons for TTS and ASR. PF, DT and KD are the three components of LRSpeech, where PF represents pre-training and fine-tuning, DT represents dual transformation, KD represents knowledge distillation. GT is the ground-truth and GT (Parallel WaveGAN) is the audio generated with Parallel WaveGAN from the ground-truth mel-spectrogram. Baseline \#1 and \#2 are two baseline methods with limited paired data.}
\label{main_results}
\vspace{-6mm}
\end{table}


\begin{figure*}[tbh] 
\centering
\hspace{-5mm}\includegraphics[width=\textwidth]{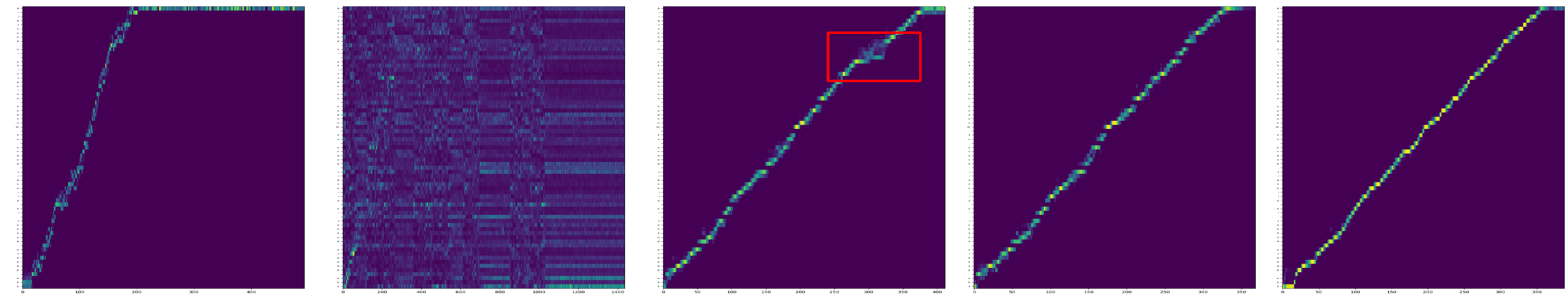}
\begin{tabular}{lllll}
  \hspace{0.4cm}(a) Baseline \#1  &
  \hspace{1.45cm}(b) Baseline \#2  & 
  \hspace{1.8cm}(c) + PF &
  \hspace{1.7cm}(d) + PF + DT & 
  \hspace{1.1cm}(e) + PF + DT + KD
\end{tabular}
\vspace{-2mm}
\caption{The TTS attention alignments (where the column and row represent the source text and target speech respectively) of an example chosen from the test set. The source text is ``the paper's author is alistair evans of monash university in australia''.} 
\label{fig_attention}
\vspace{0mm}
\end{figure*}

\subsubsection{Main Results}
We compare LRSpeech with the baselines that purely leverage the limited paired data for training, including 1) Baseline \#1, which trains TTS and ASR model only with corpus $D_h$, and 2) Baseline \#2, which adds additional corpus $D_l$ on Baseline \#1 for TTS and ASR model training. We also conduct experiments to analyze the effectiveness of each component (pre-training and fine-tuning, dual transformation, knowledge distillation) in LRSpeech.
The results are shown in Table~\ref{main_results}. We have several observations:
\begin{itemize}[leftmargin=*]
    \item Both baselines cannot synthesize reasonable speech and the corresponding IR and MOS are marked as ``/''. The WER and CER on ASR are also larger than 100\%\footnote{The WER and CER can be larger than 100\%, and the detailed reasons can be founded in Section \ref{repro:wer_cer}.}, which demonstrates the poor quality when only using the limited paired data $D_h$ and $D_l$ for TTS and ASR training. 
    \item Based on Baseline \#2, pre-training and fine-tuning (PF) can achieve an IR score of 93.09\% and a MOS score of 2.84 for TTS, and reduce the WER to 103.70\% and CER to 69.53\%, which demonstrates the effectiveness of cross-lingual pre-training for TTS and ASR.
    \item However, the paired data in both rich-resource and low-resource languages cannot guarantee high accuracy, and thus we further leverage the unpaired speech corpus $Y^u_{\text{seen}}$ and $Y^u_{\text{unseen}}$, and unpaired text corpus $X^u$ through dual transformation (DT). DT can greatly improve IR to 96.70\% and MOS to 3.28 on TTS, as well as WER to 38.94\% and CER to 19.99\%. The unpaired text and speech samples can cover more words and pronunciations, as well as more speech prosody, which help the synthesized speech in TTS achieves higher intelligibility (IR) and naturalness (MOS), and also help ASR achieves better WER and CER. 
    \item Furthermore, adding knowledge distillation (KD) brings 1.38\% IR, 0.29 MOS, 10.12\% WER and 5.34\% CER improvements. We also list the speech quality in terms of MOS for the ground-truth recordings (GT) and the synthesized speech from the ground-truth mel-spectrogram by Parallel WaveGAN vocoder (GT (Parallel WaveGAN)) in Table~\ref{main_results} as the upper bounds for references. It can be seen that LRSpeech achieves a MOS score of 3.57, with a gap to the ground-truth recordings less than 0.5, demonstrating the high quality of the synthesized speech.  
    \item There are also some related works focusing on low-resource TTS and ASR, such as Speech Chain~\cite{tjandra2017listening}, Almost Unsup~\cite{ren2019almost}, and SeqRQ-AE~\cite{liu2019towards}. However, these methods require much data resource to build systems and thus cannot achieve reasonable accuracy in the extremely low-resource setting. For example,~\cite{ren2019almost} requires a pronunciation lexicon to convert the character sequence into phoneme sequence, and dozens of hours of single-speaker high-quality unpaired speech data to improve the accuracy, which are costly and not available in the extremely low-resource setting. As a result, ~\cite{ren2019almost} cannot synthesize reasonable speech in TTS and achieves high WER according to our preliminary experiments.
\end{itemize}
As a summary, LRSpeech achieves an IR score of 98.08\% and a MOS score of 3.57 for TTS with extremely low data cost, which meets the online requirements for deploying the TTS system. Besides, it also achieves a WER score of 28.82\% and a CER score of 14.65\%, which is highly competitive considering the data resource used, and shows great potential for further online deployment.

\subsubsection{Analyses on the Alignment Quality of TTS}
\begin{table}[!t]
\centering
\begin{tabular}{l|l l}
\toprule
 Setting   &   WCR & ADR (\%) \\
\midrule
~~ PF        & 0.65 & 97.85\\
~~ PF + DT   & 0.66 & 98.37\\
~~ PF + DT + KD (LRSpeech)   & 0.72 & 98.81  \\
\bottomrule
\end{tabular}
\caption{The word coverage ratio (WCR) and attention diagonal ratio (ADR) scores in TTS model under different settings.}
\label{main_results_wcr}
\vspace{-6mm}
\end{table}

\begin{figure*}[htb]
\centering
\vspace{-2mm}
\includegraphics[width=\textwidth]{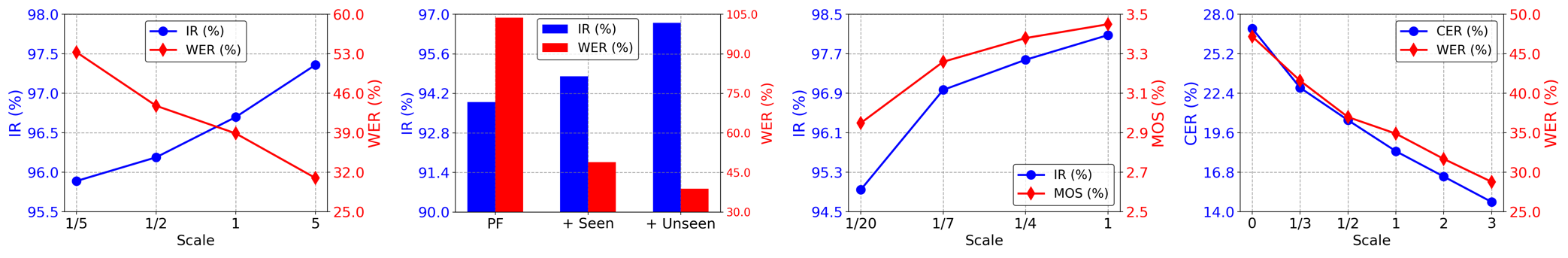}
\begin{tabular}{llll}
   \hspace{-0.2cm}\small(a) Varying the data scale of $D_l$  &
   \hspace{0.2cm}\small(b) Results using $Y^u_{\text{seen}}$ and $Y^u_{\text{unseen}}$& 
   \hspace{0.4cm}\small(c) Varying the data $X^u$ for &
   \hspace{0.8cm}\small(d)  Varying the data $X^u$ for 
   \\
   \hspace{1.2cm} &
   \hspace{2.2cm}   & 
   \hspace{0.4cm}\small  TTS knowledge distillation &
   \hspace{0.8cm}\small   ASR knowledge distillation
\end{tabular}
\vspace{-2mm}
\caption{Analyses of LRSpeech with different training data.}
\label{fig_anayses_data}
\vspace{-2mm}
\end{figure*}

Since the quality of the attention alignments between the encoder (text) and decoder (speech) are good indicators of the performance of TTS model, we analyze the word coverage ratio (WCR) and attention diagonal ratio (ADR) as described in Section~\ref{method:Distill:TTS} and show their changes among different settings in Table~\ref{main_results_wcr}. We also show the attention alignments of a sample case from each setting in Figure~\ref{fig_attention}. We have several observations: 
\begin{itemize}[leftmargin=*]
\item As can be seen from Figure~\ref{fig_attention}~(a) and (b), both Baseline \#1 and \#2 achieve poor attention alignments and their synthesized speech samples are crashed (ADR is smaller than 0.5)
. The attention weights of Baseline \#2 are almost randomly assigned and the synthesized speech is crashed, which demonstrates that simply adding a few low-quality multi-speaker data ($D_l$) on $D_h$ cannot help the TTS model but make it worse. Due to the poor alignment quality of Baseline \#1 and \#2, we do not analyze their corresponding WCR.
\item After adding pre-training and fine-tuning (PF), the attention alignments in Figure~\ref{fig_attention}~(c) become diagonal, which demonstrates the TTS model pre-training in rich-resource languages can help build reasonable alignments between text and speech in low-resource languages. Although the synthesized speech can be roughly understood by humans, it still has many issues such as word skipping and repeating. For example, the word ``in" in the red box of Figure~\ref{fig_attention}~(c) has low attention weight (WCR), and thus the speech skips the word ``in". 
\item Further adding dual transformation (DT) improves WCR and ADR, and also alleviates the words skipping and repeating issues. Accordingly, the attention alignments in Figure~\ref{fig_attention}~(d) are better. 
\item Since there still exist some word skipping and repeating issues after DT, we filter the synthesized speech according to WCR and ADR during knowledge distillation (KD). The final WCR is further improved to 0.72 and ADR is improved to 98.81\% as shown in Table~\ref{main_results_wcr}, and the attention alignments in Figure~\ref{fig_attention}~(e) are much more clear.
\end{itemize}

 

\subsection{Further Analyses of LRSpeech}
There are some questions to further investigate in LRSpeech:
\begin{itemize}[leftmargin=*]
    \item Low-quality speech data may bring noise to the TTS model. How can the accuracy change if using different scales of low-quality paired data $D_l$? 
    \item As described in Section~\ref{sec_dt_unseen}, supporting the LRSpeech training with unpaired speech data from seen and especially unseen speakers ($Y^u_{\text{seen}}$ and $Y^u_{\text{unseen}}$) is critical for a robust and scalable system. Can the accuracy be improved if using $Y^u_{\text{seen}}$ and $Y^u_{\text{unseen}}$? 
    \item How can the accuracy change if using different scales of unpaired text data $X^u$ to synthesize speech during knowledge distillation?
\end{itemize}
We conduct experimental analyses to answer these questions. For the first two questions, we simply analyze LRSpeech without knowledge distillation, and for the third question, we analyze in the knowledge distillation stage. The results are shown in Figure~\ref{fig_anayses_data}\footnote{The audio samples and complete experiments results on IR and MOS for TTS, and WER and CER for ASR can be founded in https://speechresearch.github.io/lrspeech.}. We have several observations:
\begin{itemize}[leftmargin=*]
    \item As shown in Figure~\ref{fig_anayses_data}~(a), we vary the size of $D_l$ with $1/5\times$, $1/2\times$ and $5\times$ of the default setting (1000 paired data, 3.5 hours) used in LRSpeech, and find that more low-quality paired data result in the better accuracy for TTS.
    \item As shown in Figure~\ref{fig_anayses_data}~(b), we add $Y^u_{\text{seen}}$ and $Y^u_{\text{unseen}}$ respectively, and find that both of them can boost the accuracy of TTS and ASR, which demonstrates the ability of LRSpeech to utilize unpaired speech from seen and especially unseen speakers. 
    \item As shown in Figure~\ref{fig_anayses_data}~(c), we vary the number of synthesized speech for TTS during knowledge distillation with $1/20\times$, $1/7\times$ and $1/4\times$ of the default setting (20000 synthesized speech data), and find more synthesized speech data result in better accuracy.
    \item During knowledge distillation for ASR, we use two kinds of data: 1) the realistic speech data (8050 data in total), which contains $D_h$, $D_l$ and the pseudo paired data distilled from $Y^u_{\text{seen}}$ and $Y^u_{\text{unseen}}$ by the ASR model, 2) the synthesized speech data, which are the pseudo paired data distilled from $X^u$ by the TTS model. We vary the number of synthesized speech data from $X^u$ (the second type) with $0\times, 1/3\times, 1/2\times, 1\times, 2\times, 3\times$ of the realistic speech data (the first type) in Figure~\ref{fig_anayses_data}~(d). It can be seen that increasing the ratio of synthesized speech data can achieve better results. 
\end{itemize}
All the observations above demonstrate the effectiveness and scalability of LRSpeech by leveraging more low-cost data resources.

\subsection{Apply to Truly Low-Resource Language: Lithuanian}

\paragraph{Data Setting}
The data setting in Lithuanian is similar to that in English. We select a subset of Liepa corpus~\cite{laurinvciukaite2018lithuanian} and only use the characters as the raw texts. The $D_h$ contains 50 paired text and speech data (3.7 minutes), $D_l$ contains 1000 paired text and speech data (1.29 hours), $Y^u_{\text{seen}}$ contains 4000 unpaired speech data (5.1 hours), $Y^u_{\text{unseen}}$ contains 5000 unpaired speech data (6.7 hours), and $X^u$ contains 20000 unpaired texts.

We select Lithuanian text sentences from the news-crawl dataset as the test set for TTS. We randomly select 200 sentences for IR test and 20 sentences for MOS test, following the same test configuration in English. Each audio is listened by at least 5 testers for IR test and 20 testers for MOS test, who are all native Lithuanian speakers. For ASR evaluation, we randomly select 1000 speech data (1.3 hours) with 197 speakers from Liepa corpus to measure the WER and CER scores. The test sentences and speech for TTS and ASR  do not appear in the training corpus.

\paragraph{Results}
As shown in Table~\ref{tab_exp_deploy}, the TTS model on Lithuanian achieves an IR score of 98.60\% and a MOS score of 3.65, with a MOS gap to the ground-truth recording less than 0.5, which also meets the online deployment requirement\footnote{The audio samples can be founded in https://speechresearch.github.io/lrspeech}. The ASR model achieves a CER score of 10.30\% and a WER score of 17.04\%, which shows great potential under this low-resource setting.

\begin{table}[h]
\centering
\begin{tabular}{l | l l | l l}
\toprule
Setting    &  IR (\%) & MOS & WER (\%) & CER (\%)  \\
\midrule
Lithuanian &  98.60  &  3.65  & 17.04  & 10.30 \\
\midrule
GT (Parallel WaveGAN)   &  -  &  3.89   &  -  & - \\
GT           &  -  &  4.01   &  -  & - \\
\bottomrule
\end{tabular}
\caption{The results of LRSpeech on TTS and ASR with regard to Lithuanian.}
\label{tab_exp_deploy}
\vspace{-6mm}
\end{table}

\section{Conclusion}
In this paper, we developed LRSpeech, a speech synthesis and recognition system under the extremely low-resource setting, which supports rare languages with low data costs. We proposed pre-training and fine-tuning, dual transformation and knowledge distillation in LRSpeech to leverage few paired speech and text data, and slightly more multi-speaker low-quality unpaired speech data to improve the accuracy of TTS and ASR models. Experiments on English and Lithuanian show that LRSpeech can meet the requirements of online deployment for TTS and achieve very promising results for ASR under the extremely low-resource setting, demonstrating the effectiveness of LRSpeech for rare languages.   

Currently we are deploying LRSpeech to a large commercialized cloud TTS service. In the future, we will further improve the accuracy of ASR in LRSpeech and also deploy it to this commercialized cloud service.

\section{Acknowledgements}
Jin Xu and Jian Li are supported in part by the National Natural Science Foundation of China Grant 61822203, 61772297, 61632016, 61761146003,
and the Zhongguancun Haihua Institute for Frontier Information Technology, Turing AI Institute of Nanjing and Xi'an Institute for Interdisciplinary Information Core Technology.

\bibliographystyle{ACM-Reference-Format}
\bibliography{low_tts_ref}

%
\clearpage
\appendix

\section{Reproducibility}
\label{sec:reproducibility}

\subsection{Datasets}
\label{repro:dataset}
We list the detailed information of all of the datasets used in this paper in Table~\ref{fig_dataset}. Next, we first describe the details of the data preprocessing for speech and text data, and then describe what is the ``high-quality'' and ``low-quality'' speech mentioned in this paper\footnote{We show some high-quality speech (target speaker) and low-quality speech (other speakers) from the training set in the demo page: https://speechresearch.github.io/lrspeech.}.

\paragraph{Data Proprocessing}
For the speech data, we re-sample it to 16kHZ and convert the raw waveform into mel-spectrograms following~\citet{shen2018natural} with 50ms frame size, 12.5ms hop size. For the text, we use text normalization rules to convert the irregular word into the normalized type which is easier to pronounce, e.g., ``Sep 7th" will be converted into ``September seventh". 

\paragraph{High-Quality Speech}
We use high-quality speech to refer the speech data from TTS corpus (e.g., LJSpeech, Data Baker as shown in Table~\ref{fig_dataset}), which are usually recorded in a professional recording studio with consistent characteristics such as speaking rate. Collecting high-quality speech data for TTS is typically costly~\cite{cooper2018characteristics}. 

\paragraph{Low-Quality Speech}
We use low-quality speech to refer the speech data from ASR corpus (e.g., LibriSpeech, AIShell, Liepa as shown in Table~\ref{fig_dataset}). Compared to high-quality speech, low-quality speech usually contains noise due to the recording devices (e.g., smartphones, laptops) or the recording environment (e.g., room reverberation, traffic noise). However, low-quality speech cannot be too noisy for model training.  We just use the term ``low-quality'' to differ from high-quality speech. 

\begin{table*}[h]
\small
\begin{tabular}{l|l|l|l|l|l}
\toprule
Dataset     & Type                & Speakers                                  & Language          & Open Source & Usage                                 \\
\midrule
Data Baker  & High-quality speech & Single          & Mandarin Chinese           & $\checkmark$         & Pre-training                              \\
AIShell  & Low-quality speech & Multiple             & Mandarin Chinese           & $\checkmark$         & Pre-training                                \\
LJSpeech    & High-quality speech & Single              & English           & $\checkmark$   & Training                       \\
LibriSpeech & Low-quality speech  & Multiple              & English           & $\checkmark$        & Training / Engish ASR test    \\
Liepa         & Low-quality speech  & Multiple                   & Lithuanian        & $\checkmark$       & Training / Lithuanian ASR test \\
news-crawl & Text               & /        & English/Lithuanian & $\checkmark$        & English/Lithuanian training and TTS test  \\
\bottomrule
\end{tabular}
\caption{The datasets used in this paper.}
\label{fig_dataset}
\vspace{-4mm}
\end{table*}


\subsection{Model Configurations and Training} \label{repro:model_config}
Both the TTS and ASR models use the 6-layer encoder and 6-layer decoder. For both the TTS and ASR models, the hidden size and speaker ID embedding size is 384 and the number of attention heads is 4. The kernel sizes of 1D convolution in the 2-layer convolution network are set to 9 and 1 respectively, with input/output size of 384/1536 for the first layer and 1536/384 in the second layer. For the TTS model, the input module of the decoder consists of 3 fully-connected layers. The first two fully-connected layers have 64 neurons each and the third one has 384 neurons. The ReLU non-linearity is applied to the output of every fully-connected layer. We also insert 2 dropout layers in between the 3 fully-connected layers, with dropout probability 0.5. The output module of the decoder is a fully-connected layer with 80 neurons. For the ASR model, the encoder contains 3 convolution layers. The first two are $3 \times 3$ convolution layers with stride 2 and filter size 256, and the third one with stride 1 and filter size 256. The ReLU non-linearity is applied to the output of every convolution layer except the last one.

We implement LRSpeech based on the tensor2tensor codebase\footnote{https://github.com/tensorflow/tensor2tensor}. We use the Adam optimizer with $\beta_{1}= 0.9$, $\beta_{2} = 0.98$, $\varepsilon = 10^{-9}$ and follow the same learning rate schedule in \citet{vaswani2017attention}. We train both the TTS and ASR models in LRSpeech on 4 NVIDIA V100 GPUs. Each batch contains 20,000 speech frames in total. The pre-training and fine-tuning, dual transformation and knowledge distillation take nearly 1, 7, 1 days respectively. 
We measure the TTS inference speed on a server with 12 Intel Xeon CPU, 256GB memory, 1 NVIDIA V100 GPU. The TTS model takes about 0.21s to generate 1.0s of speech, which satisfies online deployment requirements for inference speed.

\subsection{Evaluation Details} \label{sec:reproducibility:evaluation}

\begin{figure}[h] 
\small
\centering
\includegraphics[width=0.45\textwidth]{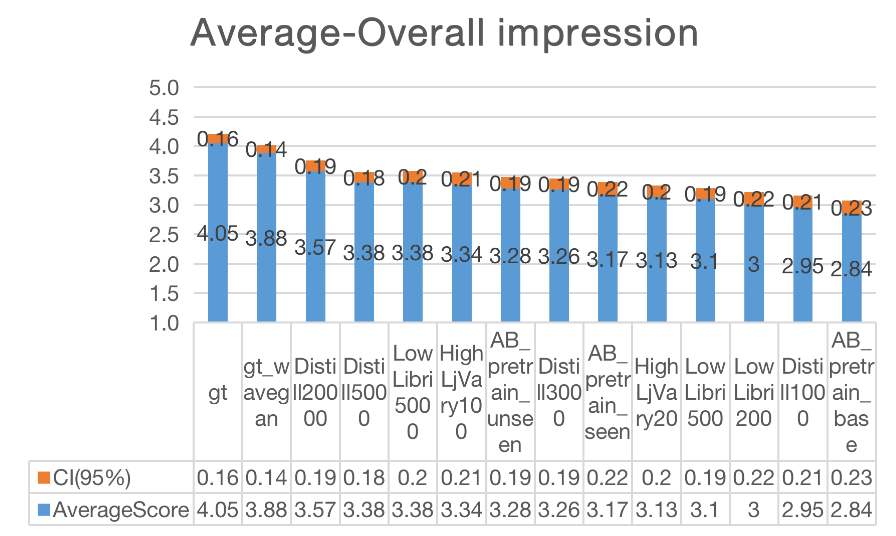}
\caption{A part of the English MOS test report.}
\label{fig_mos}
\end{figure}

\paragraph{Mean Opinion Score (MOS)} The MOS test is a speech quality test for naturalness where listeners (testers) were asked to give their opinions on the speech quality in a five-point scale MOS: 5=excellent, 4=good, 3=fair, 2=poor, 1=bad. We randomly select 20 sentences to synthesize speech for MOS test and each audio is listened by 20 testers, who are all native speakers. We present a part of the MOS test results in Figure~\ref{fig_mos}.  The complete test report can be downloaded here\footnote{https://speechresearch.github.io/lrspeech}. 

\begin{figure}[h] 
\small
\centering
\includegraphics[width=0.45\textwidth]{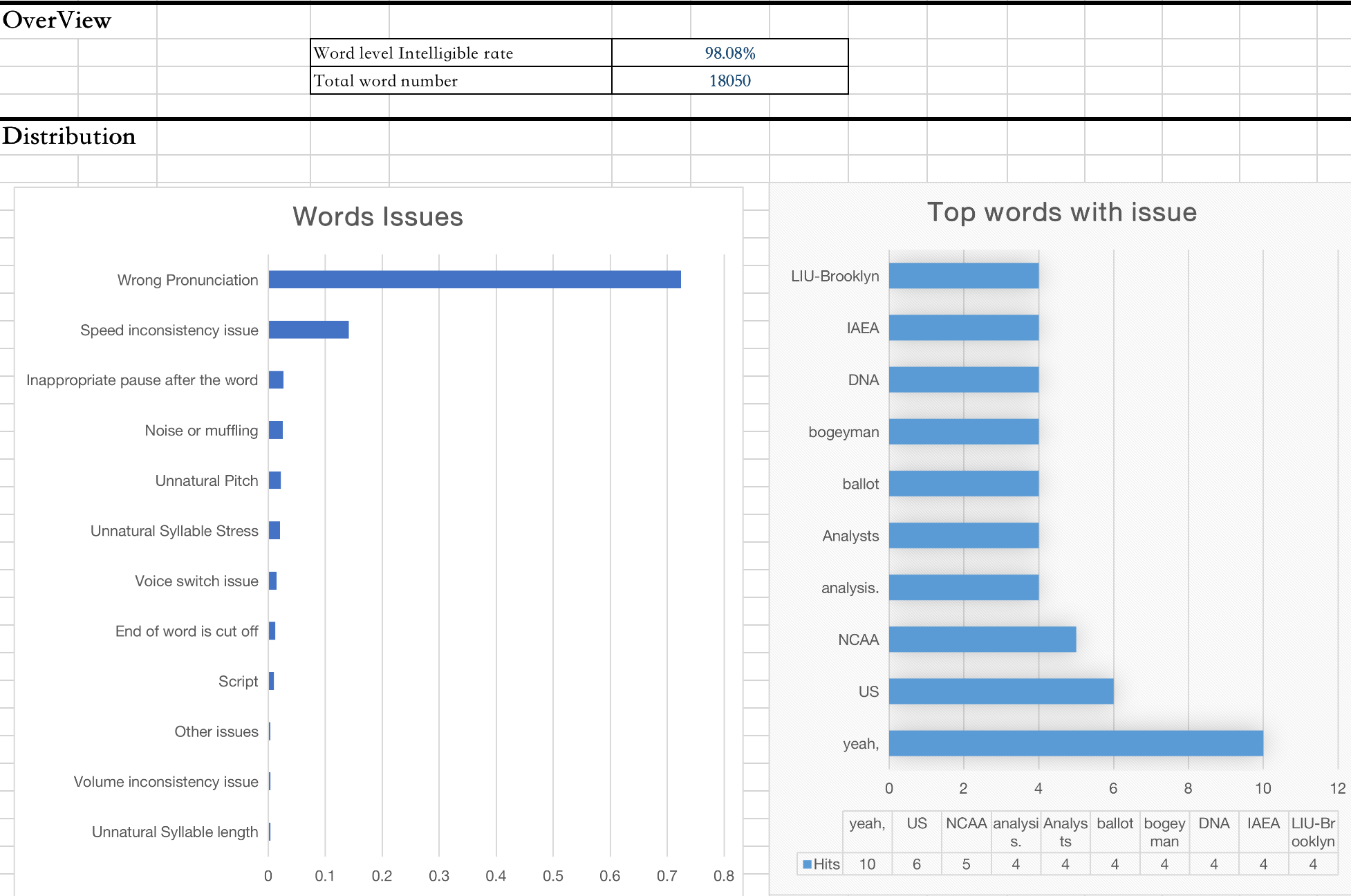}
\caption{A part of the English IR test report.}
\label{fig_IR}
\end{figure}

\paragraph{Intelligibility Rate (IR)} The IR test is a speech quality test for Intelligibility. During the test, the listeners (testers) are requested to mark every unintelligible word in the text sentence. IR is calculated by the proportion of the words that are intelligible over the total test words.  We randomly select 200 sentences to synthesize speech for IR test and each audio is listened by 5 testers, who are all native speakers. A part of the IR test results is shown in Figure~\ref{fig_IR}. You can find more test reports from the demo link. 

\begin{table*}[!ht]
\small
\centering
\begin{tabular}{l|l}
\toprule
Setting  & Result  \\
                                  \midrule
Reference  &  some mysterious force seemed to have brought about a convulsion of the elements     \\
\midrule
Baseline \#1            &  in no characters is the contrast between the ugly and vulgar illegibility of the modern type    \\
\midrule
Baseline \#2         &  the queen replied in a careless tone for instance now she went on     \\
\midrule
~~ + PF           &  some of these ceriase for seen to him to have both of a down of the old lomests       \\
\midrule
~~ + PF + DT    &   some misterious force seemed to have brought about a convulsion of the elements       \\
\midrule
~~ + PF + DT + KD (LRSpeech)  &  some mysterious force seemed to have brought about a convulsion of the elements      \\
\bottomrule
\end{tabular}
\caption{A case analysis for ASR model under different settings.}
\label{exp:asr:case_analysis}
\end{table*}

\paragraph{WER and CER} \label{repro:wer_cer} 
Given the reference text and predicted text, the WER calculates the edit distance between them and then normalizes the distance by dividing the number of words in the reference sentence. The WER is defined as $WER=\frac{S+D+I}{N}$, where the $N$ is the number of words in the reference sentence, $S$ is the number of substitutions, $D$ is the number of deletions and $I$ is the number of insertions. The WER can be larger than 100\%. For example, given the reference text ``an apple'' and predicted text ``what is history'', the predicted text needs two substitution operations and one insertion operation. For this case, the WER is $\frac{2 + 1}{2}$=150\%. The CER is similar to WER.



\subsection{Some Explorations in Experiments}
We briefly describe some other explorations in training LRSpeech in this paper:
\begin{itemize}[leftmargin=*]
\item \textbf{Pre-training and Finetune} We also try different methods such as unifying the character spaces between rich-resource and low-resource languages, or learning the mapping between the character embeddings of rich- and low-resource languages as used in \cite{chen2019end}. However, we find these methods result in similar accuracy for both TTS and ASR.

\item \textbf{Speaker Module} To design the speaker module, we explore several ways including replacing softsign with ReLU, etc. Experimental results show that the design as Figure~\ref{fig_TTS_ASR_model}~(c) can help model reduce the repeating words and missing words.


\item \textbf{Knowledge Distillation for TTS} We try to add the paired target speaker data for training. However, the result is slightly worse than that using only synthesized speech. 

\item \textbf{Knowledge Distillation for ASR} Since the synthesized speech can improve the performance, we try to remove the real speech and add plenty of synthesized speech for training. However, the ASR model cannot work well for real speech and WER is above 47\%. 

\item \textbf{Vocoder Training} In our preliminary experiments, we only use the dataset $D_{\text{rich\_tts}}$ in the rich-resource language (Mandarin Chinese) to train the Parallel WaveGAN. The vocoder can generate high-quality speech for Mandarin Chinese but fail to work for the low-resource languages. Considering that the vocoder has not been trained on the speech in the low-resource languages, we add single-speaker high-quality corpus $D_h$ and up-sample the speech data in $D_h$ to make it roughly the same with the speech data in $D_{\text{rich\_tts}}$ for training. In this way, we find that the vocoder can work well for the low-resource languages.

\end{itemize}

\subsection{Case Analyses for ASR}
We also conduct a case analysis on ASR as shown in Table~\ref{exp:asr:case_analysis}. Please refer to Section~\ref{sec_exp_result} for the descriptions of each setting in this table. The generated text by Baseline \#1 is completely irrelevant to the reference. Besides, we find from the test set that the generated text is usually the same for many completely different speech, due to the lack of paired data (only 50 paired data) for training. Baseline \#2 can generate different text sentences for different speech, but still cannot generate reasonable results. After pre-training and fine-tuning (PF), the model can recognize some words like ``have''. Similar to the effect of pre-training on TTS, pre-training ASR on rich-resource language can also help to learn the alignment between speech and text. By further leveraging unpaired speech and text, with dual transformation (DT), the generated sentence is more accurate. However, for some hard words like ``mysterious'', the model cannot recognize it correctly. TTS and ASR can also help each other not only in dual transformation but also in knowledge distillation (KD). During KD, a large amount of pseudo paired data generated from the TTS model can help the ASR model recognize most words and give correct results as shown in Table~\ref{exp:asr:case_analysis}.

\end{document}